\documentclass[]{article}
\usepackage{amssymb,epsf,rotate,amsbsy}
\newcommand{\eps}{\varepsilon}
\newcommand{\ep}{\varepsilon_6}
\newcommand{\et}{\eta}
\newcommand{\etx}{\eta^{x}}

\newcommand{\kdp}{KH$_2$PO$_4$}
\newcommand{\dekdp}{KD$_2$PO$_4$}
\newcommand{\drdp}{RbD$_2$PO$_4$}
\newcommand{\adp}{NH$_4$H$_2$PO$_4$}
\newcommand{\dadp}{ND$_4$D$_2$PO$_4$}
\newcommand{\ada}{NH$_4$H$_2$AsO$_4$}
\newcommand{\dada}{ND$_4$D$_2$AsO$_4$}

\begin{document}

Subject classification: 77.84.Fa

{\em {Institute for Condensed Matter Physics

National Academy of Sciences of Ukraine
        \footnote
	 {) 1 Svientsitskii St., L'viv-11, 79011, Ukraine.

	 \hspace{4mm}
	 Tel: +38 (0322) 707439

	 \hspace{4mm}
	 Fax: +38 (0322) 761978

	 \hspace{4mm}
         E-mail: alla@icmp.lviv.ua})
}}

{\em {
Uzhgorod State University

Physics Department
\footnote
	 {) 32 Voloshin St., Uzhgorod, UA-88000 Ukraine

	 \hspace{4mm}
	 Tel: +38 (03122) 34408

	 \hspace{4mm}
         E-mail: optics@univ.uhzgorod.ua})
}}

\vspace*{5mm}
{\bf
{
Hydrostatic pressure influence on dielectric permittivity of KH$_2$PO$_4$ and
KD$_2$PO$_4$ in the piezoelectric resonance region
}}

\vspace*{4mm}

By

\vspace*{4mm}

R. R. Levitskii$^1$, A. G. Slivka$^2$,  A. P. Moina$^1$,
P. M. Lukach$^2$, A. M. Guivan$^2$

\vspace{5mm}

\noindent
\small
We measured transverse and longitudinal dielectric
permittivities of \kdp\ and \dekdp\  in the piezoelectric resonance
frequency region under the hydrostatic pressure. The transverse
permittivity decreases with pressure in the paraelectric phase and
increases in the ferroelectric phase. The pressure dependence of
the transverse permittivity of deuterated \dekdp\ is well described by
the presented theory. The longitudinal permittivity of a pure \kdp\
exhibitsseveral peaks in the vicinity of the transition point. 
Upon increasing hydrostatic pressure, the peaks get closer and may merge.

\vspace{5mm}

\large\normalsize

\section{Introduction}
An important role in the phase transition and dielectric
response of ferroelectrics and antiferroelectrics of the \kdp\
family
is played by the geometry of hydrogen bonds in these crystals.  Pressure
dependence  of the separation $\delta$ between equilibrium proton
(deuteron) sites on a bond governs the corresponding dependences
of the transition temperature and longitudinal static dielectric
characteristics of the crystals \cite{Blinc-pressure,Torst,our!}. Moreover,
the dependence $T_{\rm C}$ on $\delta$ is universal \cite{our!,Nel5} for
several crystals of the \kdp\ type having a three dimensional network of
hydrogen bonds (MeD$_2$XO$_4$, Me = K, Rb, ND$_4$, X = P, As, \kdp, and
\adp).

High pressure studies are one of the most useful methods of exploring the
role of hydrogen subsystem  geometry  in the phase transition and
dielectric response of these crystals. Influence of hydrostatic
pressure on the transition temperature, spontaneous polarization, and Curie
constant of the  \kdp\ family ferroelectrics is well studied
experimentally. This influence for the prototype compounds \kdp\ and
\dekdp\ was theoretically described yet in Refs.
\cite{Blinc-pressure,Torst} using the four-particle cluster
approximation for the proton ordering model. The fact that this
model can account for the external pressure effects is considered
as one of its experimental  evidences.

However, much less is known about pressure effects on the transverse
dielectric characteristics of these crystals.
Only in
Ref.\cite{Kobayashi} the variation of transverse
permittivity $\eps_a$ of an undeuterated \kdp\ with hydrostatic
pressure at room temperature was measured. It was established that
$\eps_a$ decreases with pressure, and the slope $d\eps_a/d p$
decreases as frequency is increased from 400~Hz up to 25~kHz.  As far as
we know, there is no experimental data for influence of external pressures
on transverse dielectric properties
of deuterated \kdp\ type crystals. Neither it is known in what way
external pressure affects  the longitudinal or transverse dynamical
dielectric characteristics of these crystals.

The aim of the present study is to fill this gap, namely, to
explore the dependences of transverse and longitudinal dynamic
dielectric permittivities of  the \kdp\ and \dekdp\ on hydrostatic
pressure in a wide temperature range. Measurements are carried out at
the frequency of 1MHz, that belongs to the piezoelectric resonance
frequency region  of these crystals \cite{Mason_old}.  This fact should
not affect the behavior of the transverse permittivity, provided the
samples are oriented precisely, but the temperature curves of the
longitudinal permittivity must be essentially different from the
static ($0-10^4$~Hz) and high-frequency ($10^9-10^{12}$~Hz) ones.
The pressure dependence of the transverse permittivity of a deuterated
\dekdp\ is described within a microscopic theory. Thence, we can find out
whether the  pressure changes in the hydrogen bonds geometry determine
the pressure dependence of  not only the longitudinal
 \cite{Blinc-pressure,Torst,our!}, but also of the transverse dielectric
response of this crystal.

\section{Experimental methodics}
We measured the dielectric permittivity of two crystals: a pure \kdp\
(transition temperature at ambient pressure $T_{\rm C0}\sim 123$~K) and a
highly deuterated sample with $T_{\rm C0}\sim 210$~K (0.87 nominal
deuteration), hereafter abbreviated as \dekdp.  The obtained decrease of
the transition temperatures with  hydrostatic pressure $\partial T_{\rm
C}/\partial p$ is $-4.6$~K/kbar in \kdp\ and $-2.1$~K/kbar in \dekdp, that
well accords with the literature data \cite{Sam1}.

The permittivity $\eps$ was determined from the samples capacity
using the well-known formula
\[
\eps=\frac{d}{\eps_0S}C,
\]
where $d$ is the sample thickness;  $C$ is its electric capacity; $S$ is
the area of electric contacts; $\eps_0=8.85\cdot 10^{-12}$~F/m.
Capacity  was measured by the conventional bridge method
using the ac bridges E7-12 (working frequency 1~MHz) and P5016 (10~kHz
and 50kHz) within  0.2-0.4\%. As electric contacts a silver paste was
used.

Optic grade samples were placed in a autonomous hydrostatic
pressure chamber, with silicone oil serving as a pressure transmitting
medium.
Pressure was measured  by mechanical and manganin manometers with
an accuracy $\pm2$~MPa. Temperature was measured to $\pm0.1$~K with a
copper-constantan thermocouple. Capacities were obtained in a
dynamical regime with a temperature change rate $2 \cdot 10^{-2}$K/c.

\section{Experimental results}

\begin{figure}[hbt]
\begin{center}
\leavevmode
\epsfxsize=0.4\textwidth
\rotate[r]{\epsffile{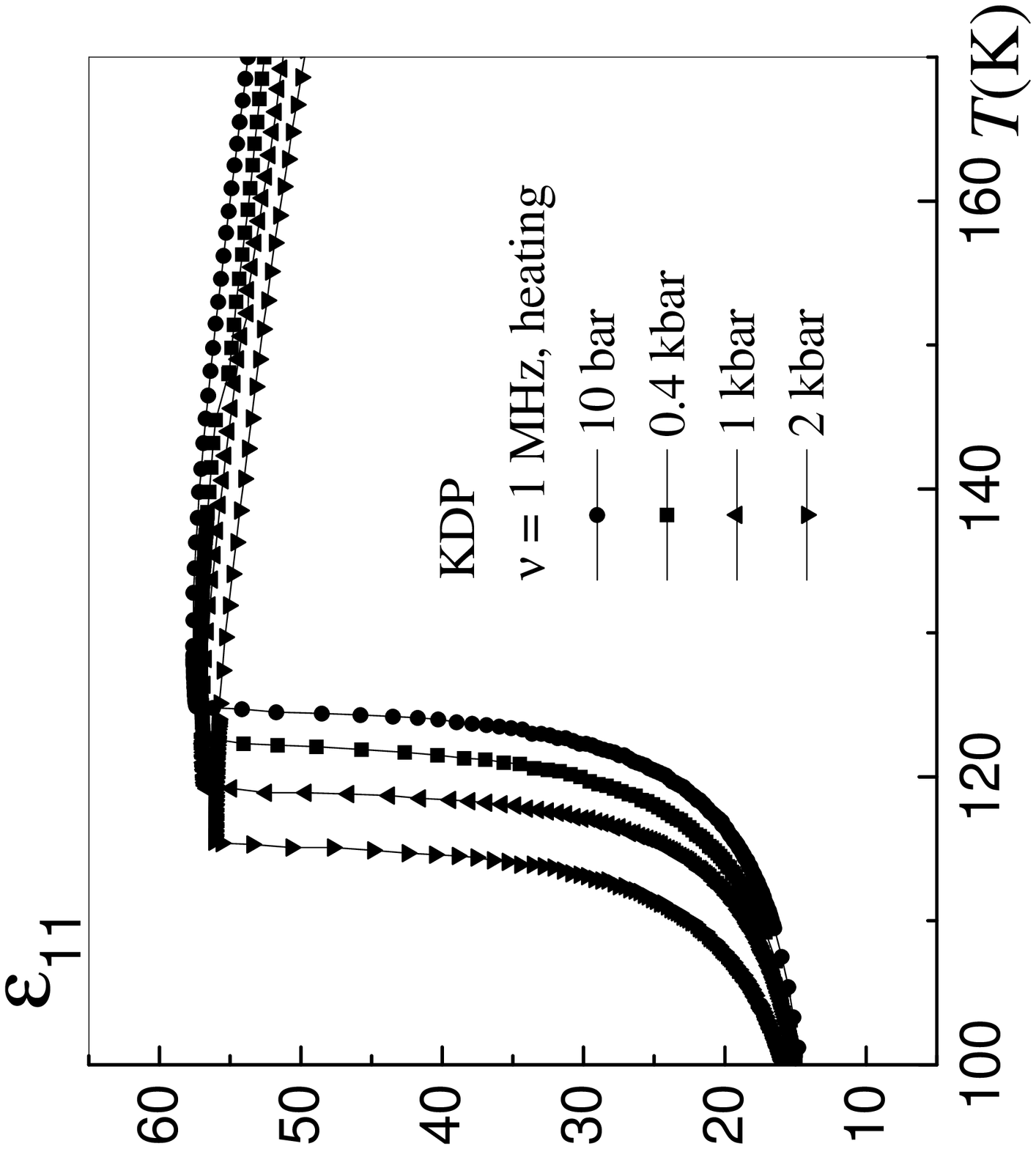}}
\hspace{1cm}
\epsfxsize=0.4\textwidth
\rotate[r]{\epsffile{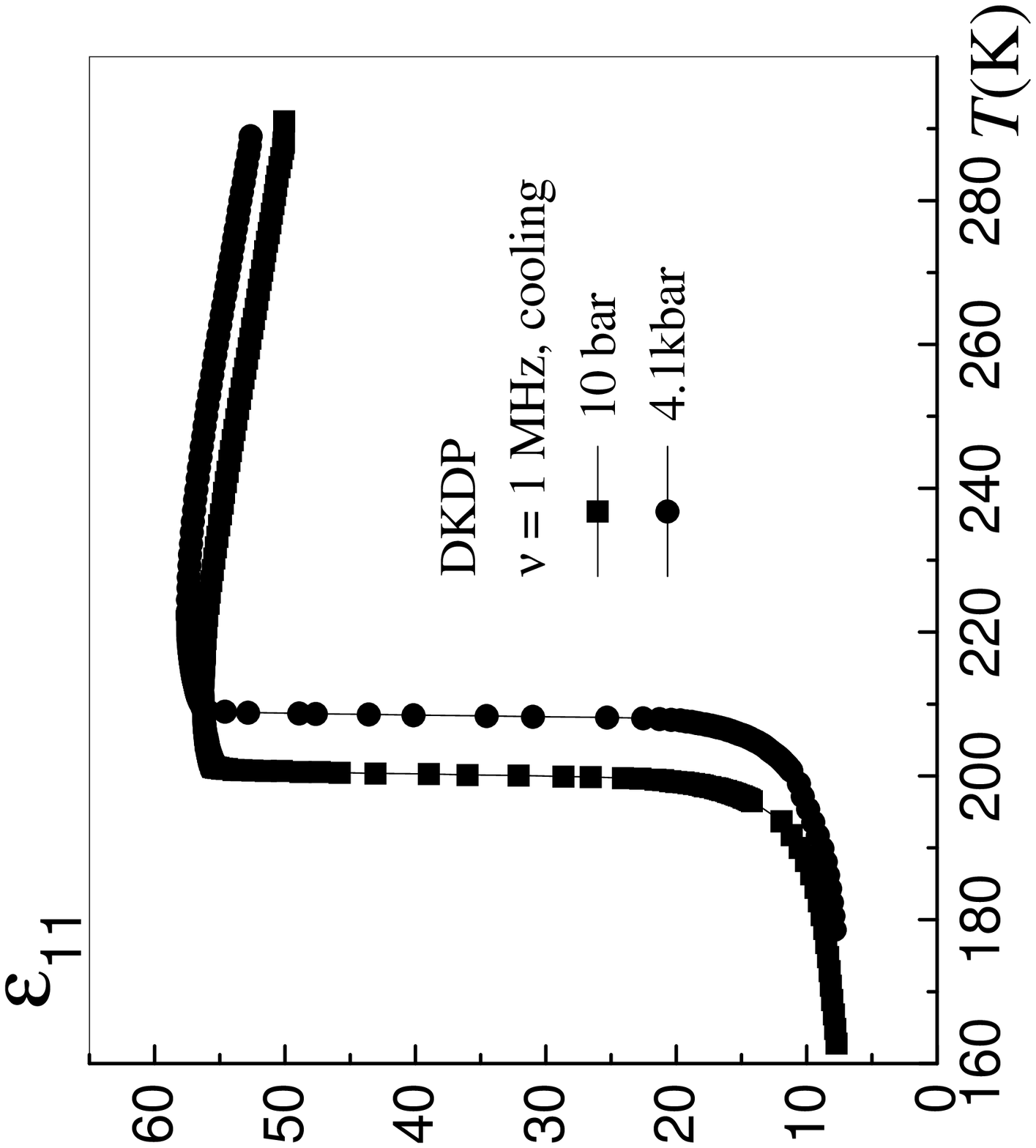}}
\end{center}
\caption{\small Transverse dielectric permittivities of \kdp\ and \dekdp\
as functions of temperature at different values of hydrostatic pressure.
The frequency is $\nu=1$MHz. Symbols are experimental points, lines
are drawn for clarity.}
\label{kdp-dkdp-a}
\end{figure}

Closeness of the measuring frequency to the piezoelectric resonance
region, as expected, does not essentially influences the transverse
permittivities of the \kdp\ and \dekdp\ ferroelectrics. The measured
curves have a typical form of the static transverse permittivity in
these crystals (see fig.~\ref{kdp-dkdp-a}).
In the paraelectric phase $\eps_{11}$ slowly increases on lowering
temperature, reaching its maximal value at  $T\approx T_{\rm  C}+15$~K,
somewhat decreases in the interval $T_{\rm  C}<T< T_{\rm  C}+15$~K,
has a jump at the transition and gradually decreases to a certain
limiting value as temperature is lowered down in the ferroelectric phase.
This gradual decrease in a deuterated \dekdp\ is much faster than in a pure
\kdp, reflecting the fact that in \kdp\ the phase transition is close to the
second order, while in \dekdp\ a pronounced first order phase
transition takes place with a significant jump of spontaneous polarization
\cite{Sam_deute}.

The obtained pressure dependences of the transverse permittivities of
ferroelectric \kdp\ and \dekdp\ are analogous to the corresponding
dependences of the transverse permittivity in antiferroelectric
\ada\ and \dada\ \cite{Gesi2}.
Under hydrostatic pressure, the magnitudes of transverse permittivities
of both crystals at temperatures far above the transition points decrease
with $\partial\ln \eps_{11}/\partial p=-3.6\%$kbar$^{-1}$ in \kdp\ at
$T=160$~K and
$-1.13\%$kbar$^{-1}$ in \dekdp\ at $T=260$~K. In the ferroelectric phase
$\eps_{11}$ increases with pressure, but this effect results from the
changes in the distance of a given temperature point from the transition
temperature, whereas the low temperature limit of $\eps_{11}$ does not depend
on pressure.

\begin{figure}[hbt] \begin{center} \leavevmode \epsfysize=0.9\textwidth
\rotate[r]{\epsffile{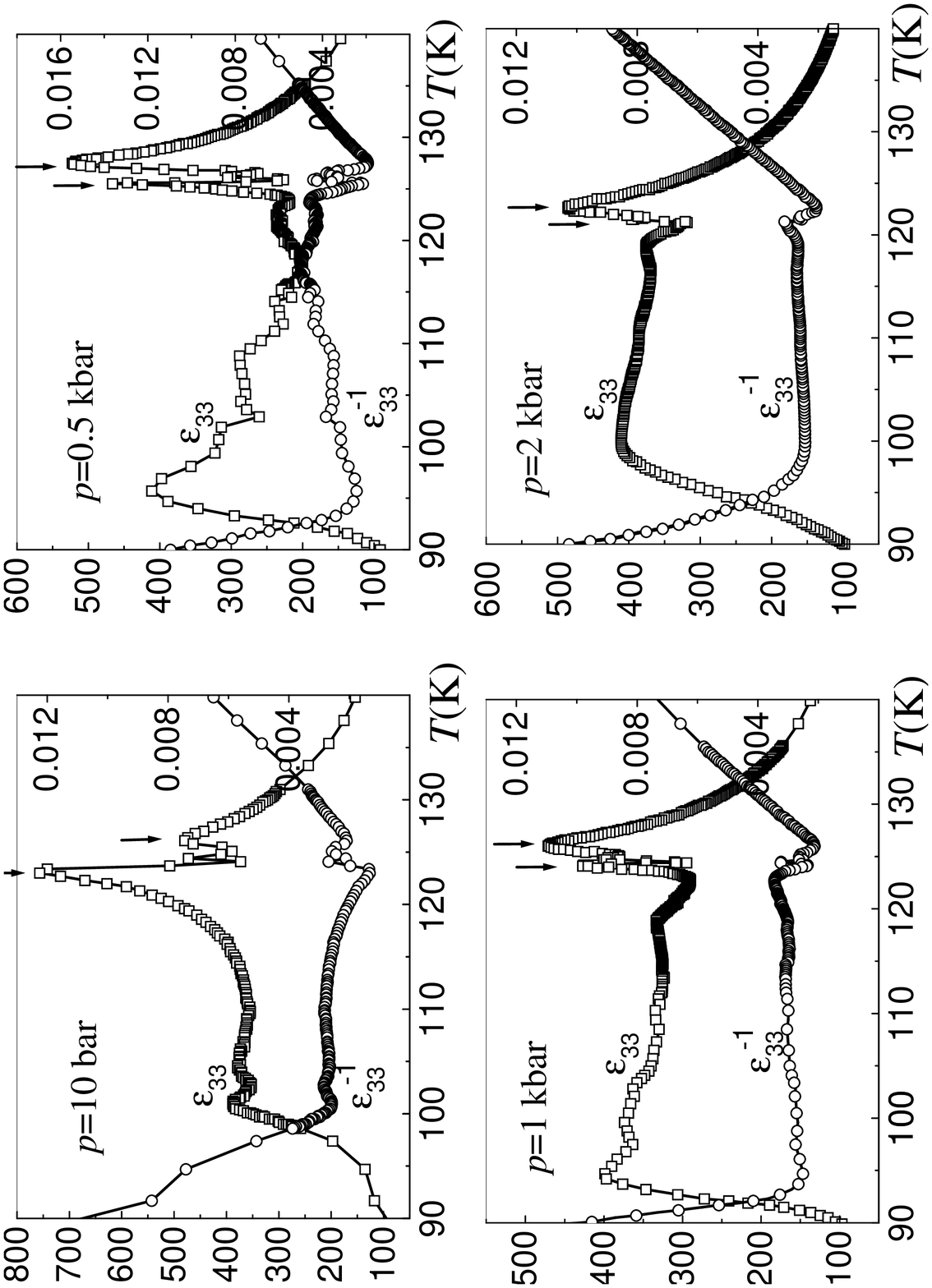}}
\end{center}
\caption{\small Longitudinal dielectric permittivity of  \kdp\  at $\nu=1$MHz
 (cooling) as a function of temperature at different values of hydrostatic
 pressure.
 Symbols are experimental points, lines
are drawn for clarity.}
\label{kdp-c}
\end{figure}

\begin{figure}[hbt] \begin{center} \leavevmode \epsfysize=0.9\textwidth
\rotate[r]{\epsffile{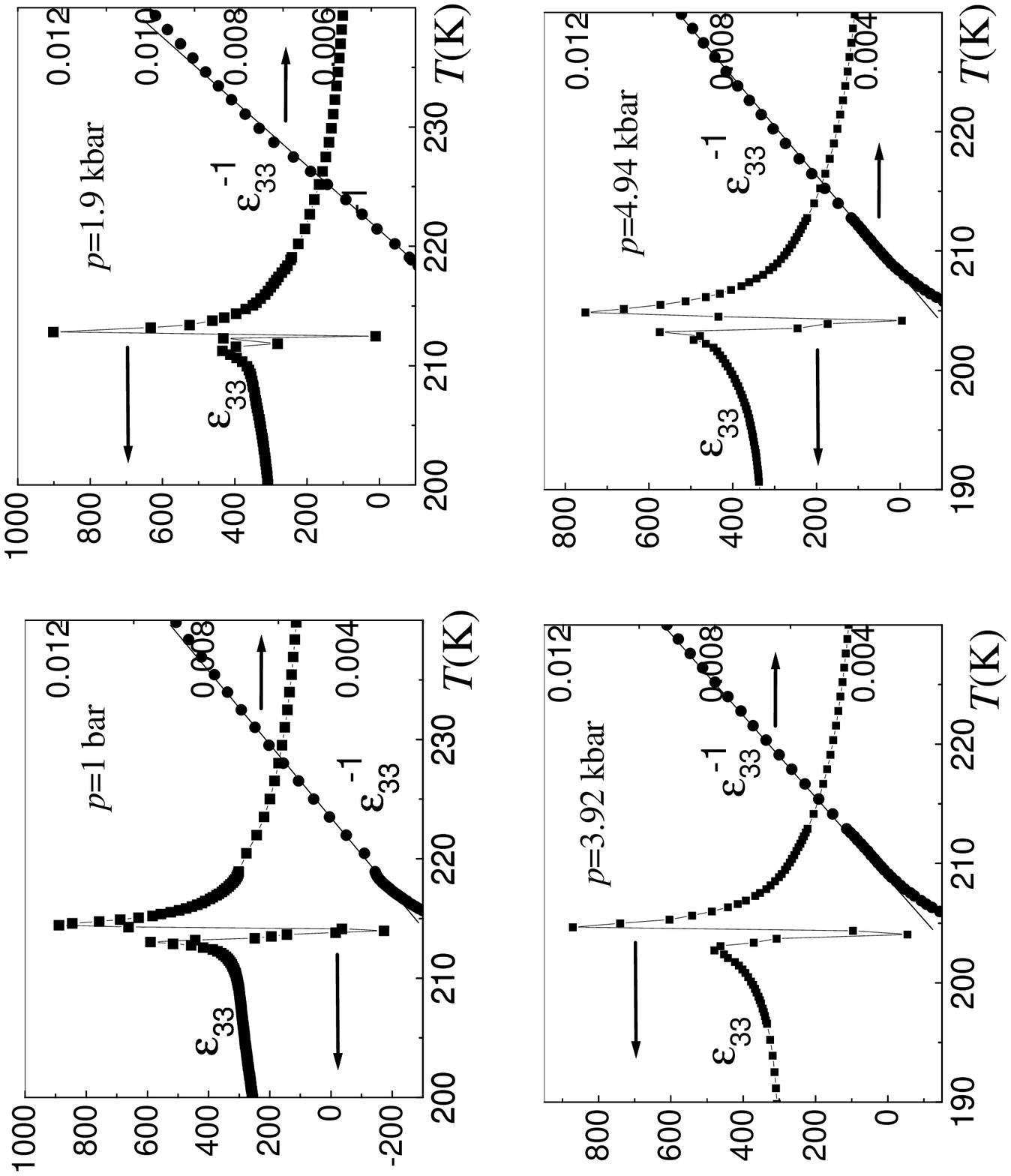}}
\end{center}
\caption{\small
Longitudinal dielectric permittivity of  \dekdp\ at $\nu=1$MHz (heating)
as a function of temperature at different values of hydrostatic pressure.
 Symbols are experimental points, lines
are drawn for clarity. The insets show the permittivity in the vicinity of
the transition point.}
\label{dkdp-c}
\end{figure}

In figs.~\ref{kdp-c} and \ref{dkdp-c} we presented the temperature curves
of the longitudinal dielectric permittivity of a pure \kdp\ and
undeuterated \dekdp, respectively, at different
values of hydrostatic pressure. Note an interesting behavior of the
permittivity of \kdp, which  exhibits  three clear maxima
(correspondingly, two minima) in the vicinity of the transition point.
Possibly due to the narrowness of the temperature interval in which
the peaks of the longitudinal permittivity occur, in a deuterated
\dekdp\ the three-maximum structure was detected only at
$p=1.9$~kbar (see  fig.~\ref{dkdp-c}). For the other pressures, only one
minimum of the permittivity was clearly observed. However, the bends in the
curves of $\eps_{33}$ in the temperature interval between the detected
paraelectric and ferroelectric peaks indicate that the second,
ferroelectric phase minimum must occur in \dekdp\ as well.

\begin{figure}[hbt] \begin{center} \leavevmode \epsfxsize=0.6\textwidth
\epsffile{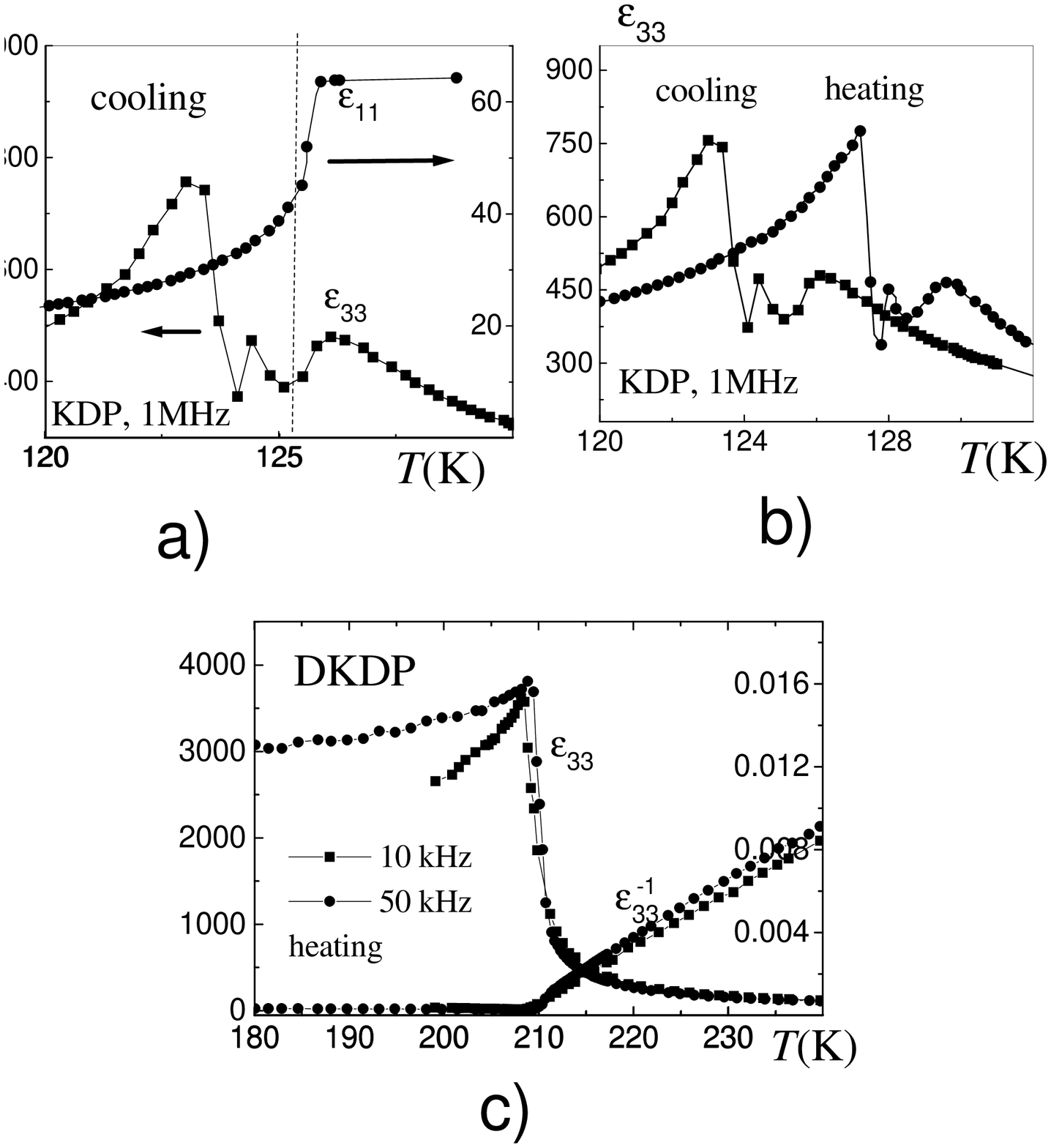}
\end{center}
\caption{\small Dielectric permittivities of \kdp\  and \dekdp\ as
functions of temperature at ambient pressure. Measuring frequencies and
direction of temperature changes (heating or cooling) are indicated at
figures. Symbols are experimental points, lines
are drawn for clarity.}
\label{kdp-c-add}
\end{figure}

As follows from fig.~\ref{kdp-c-add}a, where the curves of longitudinal
and transverse permittivities $\eps_{33}$ and $\eps_{11}$ are compared,
the right maximum occurs at the transition point. The two other maxima and
the left minimum, which take place in the ferroelectric phase, are
reproduced at different values of hydrostatic pressure as well as at
heating or cooling (see~fig.~\ref{kdp-c-add}b). At lower frequencies
($\nu=10$~kHz and $\nu=50$~kHz), the  $\eps_{33}$  of deuterated \dekdp\
has a typical form of the static permittivity with one peak at the
transition point and the Curie-Weiss behavior in the paraelectric phase
(fig.~\ref{kdp-c-add}c).

Under hydrostatic pressure, the magnitude of the  transition point maximum
in \kdp\ does not change, while that of the left ferroelectric one is
reduced from $\sim770$ at 10bar down to $\sim400$ at  $p=2$~kbar. Beside,
the temperature interval where the extrema of the permittivity occur
becomes narrower, so that the two-minimum structure of $\eps_{33}$
transforms to a single-minimum one.

At temperatures below the peaks, the longitudinal permittivity of
deuterated \dekdp\ has a typical shoulder-like form, caused by
the multi-domain structure  of the samples. Qualitatively similar
dependences can be traced in a pure \kdp\ as well, but the ``fine
structure'' of  $\eps_{33}$ here is fancier. Hydrostatic pressure almost
does not affect the shape of the temperature curve  $\eps_{33}(T<T_{\rm
C})$. Width of the plateau is about 30~K in both crystals, with the height
of about 400 being practically pressure independent.

As seen in figs.~\ref{kdp-c} and \ref{dkdp-c}, the Curie-Weiss law is
obeyed for $\eps_{33}^{-1}$  of \kdp\ and \dekdp\ above the
transition point in a  rather wide temperature range, except for a
vicinity of the transition point. The Curie constants are very close to the
corresponding static ones
\cite{Sam1} and decrease with pressure linearly from $C=2949$~K in \kdp\
and $C=3770$~K in \kdp\ at ambient pressure
with the slopes $\partial\ln C/\partial p=-0.48\%$kbar$^{-1}$ in \kdp\ and
$-1.5\%$kbar$^{-1}$ in \dekdp\ (the literature data for the Curie constants
of static permittivities are $C=2910$~K and $\partial\ln C/\partial
p=-0.66\%$kbar$^{-1}$ in \kdp\ and
3700~K and $-1.43\%$kbar$^{-1}$ in \dekdp, respectively \cite{Sam1}).

Note a strong dependence of the logarithmic pressure derivatives of the
dielectric characteristics of the crystals on deuteration and the fact
that this dependence is different for the transverse and longitudinal
quantities: $|\partial\ln \eps_{11}/\partial p|$ in \kdp\ is much higher
than in \dekdp, whereas for $|\partial\ln C/\partial p|$ the reverse holds.
A similar effect has been revealed earlier \cite{Sam1} for the transition
temperature, saturation polarization and Curie constant of these crystals.

\section{Theory}
\label{theory}
We restrict our theoretical calculations by the case
of highly deuterated
\dekdp. The known models of dielectric relaxation in hydrogen-bonded
crystals \cite{Yoshimitsu,lzm4}, based on Glauber's \cite{Glaub}
dynamics of pseudospins, are suitable  for description of the
high-frequency permittivities ($10^9-10^{12}$~Hz) only, when a crystal is
effectively clamped \cite{Kanzig}. Therefore, in order to describe the
above presented pressure dependences of the longitudinal dielectric
permittivity one must
develop a special  model of dielectric relaxation which would take into
account dynamics of piezoelectric strain $\ep$ as well. However, since
this dynamics is not essential for the transverse dielectric
response of \dekdp, the latter can be described within static models. We
shall do that within  the proton ordering model modified to the case of
strained crystals \cite{our!}.

Calculations are performed with the conventional  four-particle cluster
Hamiltonian
\begin{eqnarray}
\label{Hamiltonian}
&&  {H}_{q}= V\left[\frac{\sigma_{q1}}{2}\frac{\sigma_{q2}}{2}+
       \frac{\sigma_{q2}}{2}\frac{\sigma_{q3}}{2}+
       \frac{\sigma_{q3}}{2}\frac{\sigma_{q4}}{2}+
       \frac{\sigma_{q4}}{2}\frac{\sigma_{q1}}{2}
 \right]  \nonumber\\
&& +  U\left[
       \frac{\sigma_{q1}}{2}\frac{\sigma_{q3}}{2}+
       \frac{\sigma_{q2}}{2}\frac{\sigma_{q4}}{2}
\right]
+\Phi
 \frac{\sigma_{q1}}{2}\frac{\sigma_{q2}}{2}
 \frac{\sigma_{q3}}{2}\frac{\sigma_{q4}}{2}-
\sum_{f=1}^4\frac{z_{qf}}{\beta}
   \frac{\sigma_{qf}}{2}.
\end{eqnarray}
Two eigenvalues of the Ising pseudospin
$\sigma_{qf}=\pm 1$ are assigned to two equilibrium deuteron sites on the
$f$-th bond in the $q$-primitive cell.
The internal fields $z_{qf}$ include a long-range interactions between
deuterons (dipole-dipole and lattice mediated) taken into account in the
mean field approximation, cluster fields $\Delta_{qf}$ which describe the
short-range interactions with the pseudospin
$\sigma_{qf}$ not explicitly included into the cluster Hamiltonian, and an
external electric field applied along the  $a$-axis of the tetragonal unit
cell
\begin{equation}
\label{3:longitud}
z_{qf}
=\beta\left[-\Delta_{qf}+\sum_{q'f'}J_{ff'}(qq')
\frac{\langle\sigma_{q'f'}\rangle}{2}+\mu_{qf}^1E_1\right],
\end{equation}
$\mu_{qf}^1$ is the projection on the $a$-axis of the effective dipole
moment created by displacements of heavy ions and redistribution of electron
density induced by deuteron ordering. In the case of transverse electric
field $E_1$, the following relations
between the mean values of pseudospins and between the effective dipole
moments $\mu_{qf}^1$ are obeyed
\begin{eqnarray*}
\label{1:e12a}
&&\etx_{24}=\langle\sigma_{q2}\rangle=\langle\sigma_{q4}\rangle,\qquad
z_{24}=z_{q2}=z_{q4};\\
&& \mu_1=\mu_{q1}^1=-\mu_{q3}^1, \qquad
\mu_{q2}^1=\mu_{q4}^1=0. \nonumber \end{eqnarray*}
Parameters of the short-range
deuteron correlations $U$, $V$, $\Phi$ are
functions of the so-called Slater energies $\eps$, $w$, $w_1$:
\begin{equation}
 V=-\frac{w_{1}}{2}, \qquad
 U=-\varepsilon+\frac{w_{1}}{2} ,\qquad
 \Phi=4\varepsilon-8w+2w_{1}.
\end{equation}
Pressure dependences of the Slater energies and parameters of the
long-range interactions  are modelled as \cite{our!}
\begin{equation}
\label{Slater1}
 \varepsilon=\varepsilon^0
\Big[1-\frac 2S\frac{\delta_1}{\delta_0}
\sum_{j=1}^3\eps_j\Big],\quad
w=w^0\Bigg[1-\frac 2S\frac{\delta_1}{\delta_0}
\sum_{j=1}^3\eps_j\Big],\quad
 w_1=w^0_1\Big[1-\frac
2S\frac{\delta_1}{\delta_0}
\sum_{j=1}^3\eps_j\Big] .
\end{equation}
 and
\begin{equation} \label{2:longitud}
J_{ff'}(qq')=J^{(0)}_{ff'}(qq')\left[1-\frac 2S\frac{\delta_1}{\delta_0}
\sum_{j=1}^3\eps_j\right]+
         \sum_{j=1}^3\psi_{ff'}^j(qq')\eps_j.
\end{equation}
Here we take into account the quadratic dependence of these parameters on
the separation  $\delta$ between two equilibrium deuteron sites on a bond,
whereas the pressure dependence of  $\delta$ is known to be linear
\cite{Nel1}
\[
\delta=\delta_0+\delta_1p=\delta_0\left(1-\frac 2S\frac{\delta_1}{\delta_0}
\sum_{j=1}^3\eps_j\right),
\]
$S=\sum_{ij=1}^3S_{ij}^{(0)}$.
According to \cite{Torst,our!} we suppose that the only essential mechanism
of the pressure influence on the Slater energies is a decrease in the
D-site distance $\delta$. However, for the parameters of the
long-range interactions there exist other important mechanisms of variation
with pressure (for instance, the dipole-dipole interactions
increase when the distance between deuterons is reduced), taken into account
in (\ref{2:longitud}) via the expansions in the diagonal
components of the strain tensor $\eps_i$ ($i=1,2,3$).

Excluding the fields $\Delta_{qf}$ from $z_{qf}$ by making use of the
condition that the mean values
$\etx_{qf}\equiv\langle\sigma_{qf}\rangle$ calculated with Hamiltonian
(\ref{Hamiltonian})
and with the one-particle Hamiltonian
\[
H_{qf}^{(1)}=
-\left(\frac{z_{qf}}{\beta}-\Delta_{qf}\right)\frac{\sigma_{qf}}{2}
\]
coincide, we obtain
\begin{eqnarray*}
&&
z_{1,3}=\frac{1}{2}\ln{\frac{1+\etx_{1,3}}{1-\etx_{1,3}}}+
   \beta\left[\nu_{1}{}\etx_{1,3}+
  \nu_{3}{}\etx_{3,1}
   +2\nu_{2}{}\etx_{24}\pm
   \frac{\mu_1E_1}{2}\right],\\
&&
z_{24}=\frac{1}{2}\ln{\frac{1+\etx_{24}}{1-\etx_{24}}}+
   \beta\left[\nu_{2}{}[\etx_1+\etx_3]
   +[\nu_{1}{}+\nu_{3}{}]\etx_{24}\right],\\
&&
\etx_{1,3}=\frac{1}{D^x}\left[\sinh A_{1}+d\sinh A_{2}\pm 2a\sinh
A_{3}+
 b(2\sinh A_{4}\pm\sinh
A_{5}\pm\sinh A_{6})\right],\\&&
\etx_{24}=\frac{1}{D^x}\left[\sinh A_{1}-d\sinh
A_{2}+ b(\sinh A_{5}-\sinh
A_{6})\right],
   \end{eqnarray*}
where the following notations are used
\begin{eqnarray*}
&& D^x=\cosh A_{1}+d\cosh A_{2} +2a\cosh
A_{3}+ b(2\cosh A_{4}+\cosh A_{5}+\cosh A_{6}),\\
&& A_{1,2}=\frac{z_1^x+z_3^x}{2}\pm z_{24}^x, \quad
 A_{3,4}=\frac{z_1^x\mp z_3^x}{2}, \quad
 A_{5,6}=\frac{z_1^x-z_3^x}{2}\pm z_{24}^x; \\
&& a_{}=\exp{(-\beta\varepsilon)},\quad
b_{}=\exp{(-\beta w)},\quad
d_{}=\exp{(-\beta w_{1})},\\
&& \nu_{j}=\frac{J_{1j}(0)}{4}
   \end{eqnarray*}
$J_{ij}(0)$ is the long-range interaction matrix Fourier transform.

From the field-induced polarization of the crystal along
the $a$-axis
\[
P_1=\sum_f \frac{\mu_{qf}^i}{v}\frac{\langle\sigma_{qf}\rangle}{2}=
\frac{2\mu_1(\et_1^x+\et_3^x)}{v},
\]
 the transverse static
dielectric permittivity
$\eps_{1}^{\varepsilon}(0,T,p)$ of a clamped crystal (at $\eps_i=$const) can
be calculated
\begin{equation}
\label{clamped}
 \left.\eps_{11}^\eps(0,T,p)=\eps_{1\infty}+
 4\pi
   \left(\frac{\partial P_{1}}{\partial
E_1}\right)_{\displaystyle\varepsilon}\right|_{E_1=0}=
\eps_{1\infty}+
 4\pi
\frac{\beta\mu^2_{1}}{v}
       \frac{2(a+b\cosh z)}
        {D-2(a+b\cosh z)\varphi},
  \end{equation}
where
\begin{eqnarray*}
 \varphi=\frac{1}{1-\eta^2}+\beta\nu_a,\quad
\nu_a=\nu_1-\nu_3=
\nu_a^{0}\left[1-\frac 2S\frac{\delta_1}{\delta_0}
\sum_{j=1}^3\eps_j\right]+
\sum_{i=1}^3\psi_{ai}\eps_i,
\end{eqnarray*}
$\eps_{1\infty}$ is a high-frequency contribution to the permittivity.
The dielectric permittivity of a free crystal (at $p$=const)
$\eps_i^F(0,T,p)$ is related to (\ref{clamped})
by
\begin{eqnarray}
\label{free}
&&\eps_{11}^F(0,T,p)= \eps_{11}^\eps(0,T,p)
+4\pi d^2_{14}c_{44}^E,
\end{eqnarray}
where $d_{14}$  is the piezomodule, and $c^E_{44}$ is the elastic constant
of a short-circuited crystal.

For the involved  quantities, the following relations hold in the zero
 external field limit
\begin{eqnarray*}
&&\et\equiv
\langle\sigma_{q1}\rangle=\langle\sigma_{q2}\rangle=
         \langle\sigma_{q3}\rangle=\langle\sigma_{q4}\rangle;\\
&&  z\equiv z_{q1}=z_{q2}=z_{q3}=z_{q4}=
\frac{1}{2}\ln{\frac{1+\et}{1-\et}}+
   \beta\nu_{c}(0)\et;\\
&& D= \cosh 2z +4b\cosh z+2a+d, \\
&&  \nu_c(0)= \frac{1}{4}
\left(J_{11}(0)+2J_{12}(0)+J_{13}(0)\right)
=
\nu_c^{0}(0)\left[1-\frac 2S\frac{\delta_1}{\delta_0}
\sum_{j=1}^3\eps_j\right]+
\sum_i\psi_{ci}(0)\eps_i.
 \end{eqnarray*}
The order parameter  $\eta$ is found by minimization of the free
energy
\begin{eqnarray}
&& f = \frac{\bar v}{2}\sum_{ij}
c_{ij} \varepsilon_{i}\varepsilon_{j} -2w+
    2\nu_c(0)\eta^2 + 2T\ln
             \frac{2}{\left(1-\eta^2\right) D},
\end{eqnarray}
 whereas to find the strains we should
solve the system of equations
\begin{equation}
-p=\sum_{j=1}^3 c_{ij}\varepsilon_j,
\end{equation}
where $c_{ij}$ are the elastic constants of the whole crystal, being
determined from an experiment. Contribution of a deuteron subsystem to
pressure or temperature dependences of the elastic constants is neglected.

\section{Numerical analysis}
The values of the theory parameters for a deuterated
crystal  \dekdp\ with the transition temperature at ambient pressure $T_{\rm
C0}=210$~K are presented in Tables \ref{parameters},\ref{elastic}. They
were found in Ref. \cite{our!} and used for a description of the
uniaxial pressure $p=-\sigma_3$ dependences of the transition temperature,
longitudinal dielectric permittivity, and spontaneous polarization of the
crystal. The elastic constants of the paraelectric crystals
coincide with the experimental data of Ref. \cite{Shuv}. The new parameters
$\nu_a$, $f_1^0=(\mu_1^{0})^2/v$, and $\eps_{1\infty}$ are found by fitting
the theoretical temperature curve of the transverse permittivity to
experimental data ($\nu_a$ sets the slope $\partial \eps_{11}/\partial T$,
whereas  $f_1$, and $\eps_{1\infty}$ give the magnitude of $\eps_{11}$.

\begin{table}[hbt]
\caption{\small The theory parameters for a \dekdp\ crystal with
$T_{\rm C0}=210$~K, $\partial T_{{\rm C}}/\partial p =- 2.1
$~K/kbar. Plus and minus indices denote the quantities used
in the paraelectric and ferroelectric phases, respectively.}
\begin{center}
\begin{tabular}{cccccccccccc}
\hline
 $ \eps^0$
& $ w^0$
& $ \nu^0_c(0)$
& $ \nu^0_a(0)$
& $f^{0+}_1$
& $f^{0-}_1$
& $\psi_{c1}^-$
& $\psi_{c2}^-$
& $\psi_{c3}^-$
& $\psi_{c1}^+$
& $\psi_{c3}^+ $
&${\delta_1}/{\delta_0}\cdot10^{3}$\\
\multicolumn{11}{c}{(K)} &
  (kbar$^{-1}$)
 \\
\hline
 87.6 & 785 & 37.05 & -32 & 830 &  520 & 120 & 100  & $-$545       &
        110 &        $-$545
    & $- 7.5$ \\
\hline
\end{tabular}

\end{center}
\label{parameters}
\end{table}

\begin{table}
\caption{\small Elastic constants
 of the considered crystal (units of $10^5$bar).}
 \begin{center}
\begin{tabular}{@{}cccccccccc}
\hline
$c_{11}^{+}$ & $ c_{12}^{+}$ & $c_{13}^{+}$ & $c_{33}^{+}$
    & $c_{11}^{-}$ & $ c_{12}^{-}$ & $c_{13}^{-}$ &
      $c_{22}^{-}$ & $ c_{23}^{-}$ & $c_{33}^{-}$ \\
      \hline
       6.93& $-$0.78& 1.22 & 5.45
     & 6.8    & $-$0.78  &  1.0 & 6.99  & 1.0  & 5.3\\
     \hline
\end{tabular}
\end{center}
 \label{elastic}
 \end{table}

The so-called deformation potentials
$\psi_{ai}$, which enter only the expression for the transverse
permittivity, do not essentially affect the permittivity. Therefore,
for the sake of simplicity we put  $\psi_{ai}=0$. Contributions of
the double-ionized deuteron configurations (with four deuterons
close to a given PO$_4$ group or with none) are neglected as well by
taking $w_1\to\infty$.

Hence, the only new theory parameter, governing the pressure
dependence of the calculated characteristics, is the ratio
$\delta_1/\delta_0$ -- the relative rate of the
pressure changes in the D-site separation $\delta$. We choose it so
that at all other parameters unchanged, the correct dependence of
the transition temperature on the hydrostatic pressure $\partial T_{{\rm
C}}/\partial p =- 2.1 $~K/kbar is obtained.

In Ref. \cite{our!} we have shown that at the chosen analogously values of
$\delta_1/\delta_0$, the recalculated dependences
$T_{\rm C}(\delta)$  for six deuterated ferroelectric and antiferroelectric
crystals with a three dimensional network of hydrogen
bonds MeD$_2$XO$_4$ (Me = K, Rb, ND$_4$, X = P, As) form a single universal
linear dependence, while $T_{\rm C}$ and $\delta$ can be  altered  by either
hydrostatic or uniaxial $p=-\sigma_3$ (for
KD$_2$PO$_4$) pressures. Experimentally this universality has been
established  by R. Nelmes  {\it et al} \cite{Nel5} for undeuterated \kdp\ and
\adp\ and deuterated  \dekdp\ and \dadp. At  the adopted
in this paper value of  $\delta_1/\delta_0$ for
\dekdp\ with $T_{\rm C0}=210$~K, the dependence $T_{\rm
C}(p)[\delta(p)]$ for this specific crystal also accords with the universal
line \cite{our!}.

The slopes $\partial\mu_i/\partial p$  can be determined without
introducing into the theory any extra fitting parameter on the basis of the
following simple half-empirical speculations, nevertheless
yielding a fair  agreement with the experiment.

It is believed that the deuteron ordering in the system
results in displacements of heavy ions and electron density which
contribute to crystal polarization.  Since, when ordered, a deuteron shifts
from its central position on a hydrogen bond to the off-central one by a
distance $\delta/2$, according to \cite{our!} we assume that the heavy ions
displacements are also proportional to $\delta$ and that
$\mu_i$ is proportional to the corresponding lattice constant $a_i$. This
yields
\begin{equation}
\label{effective}
\frac{1}{\mu_i^0}\frac{\partial\mu_i}{\partial p}=
\frac{\delta_1}{\delta_0}+\frac{\eps_i}{p}.
\end{equation}
The pressure dependence of the effective longitudinal dipole
moment $\mu_3$, calculated with (\ref{effective}) provides a fair agreement
with the available  experimental data for the pressure dependence of
spontaneous polarization of \dekdp\ and of the static dielectric
permittivities of \dekdp\ and \drdp\ \cite{our!}.

\begin{figure}[hbt]
\begin{center}
\leavevmode
\epsfxsize=6.5cm
\rotate[r]{\epsffile{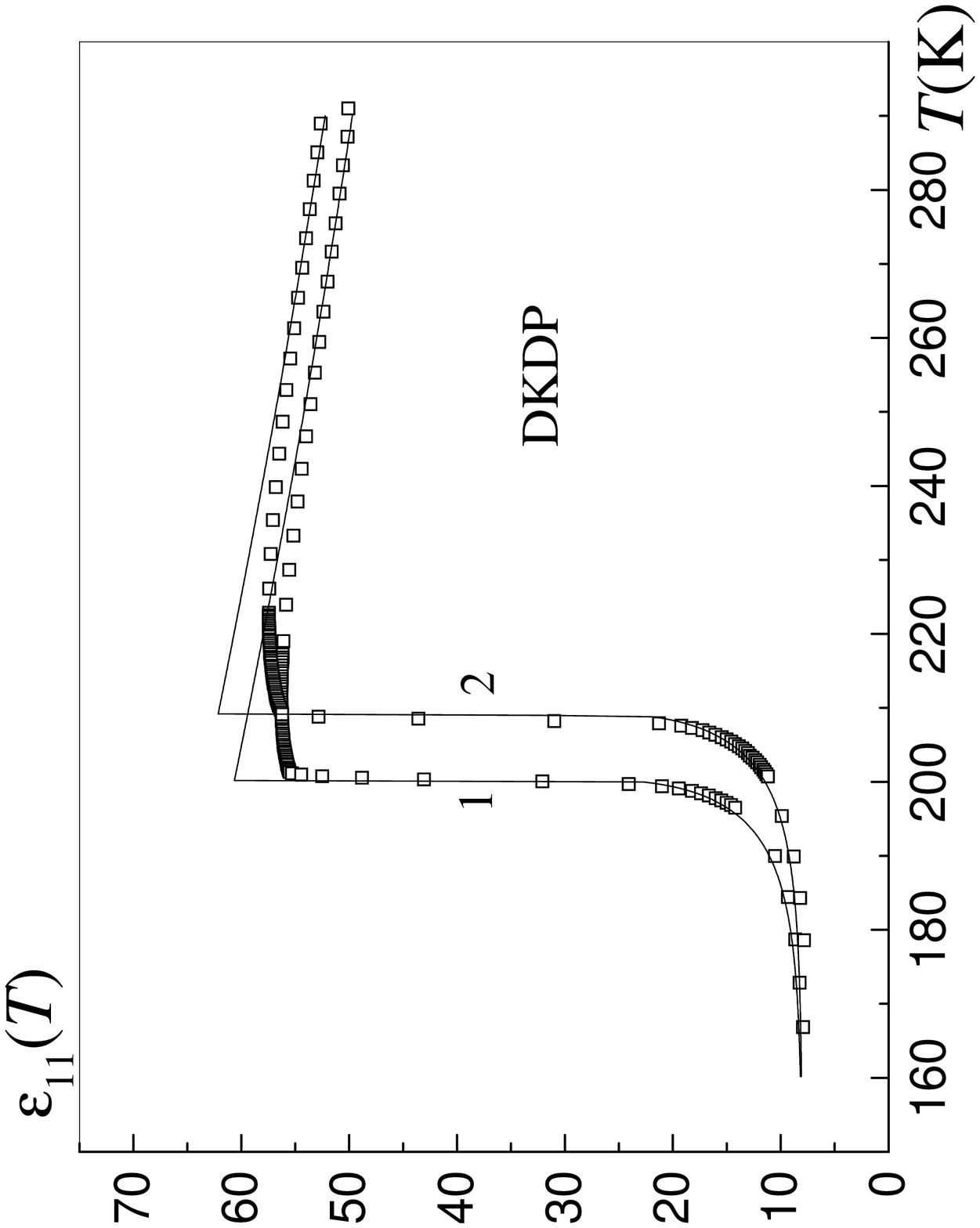}}
\end{center}
\caption{\small Transverse dielectric permittivity of \dekdp\ as a function
of temperature at different values of hydrostatic pressure (kbar)
1 -- 0;
2 -- 4.1. Lines are calculated theoretically; symbols are experimental
points of the present work. }
\label{transverse}
\end{figure}

In fig.~\ref{transverse} we depict the theoretical curves of the
transverse static dielectric permittivity of
\dekdp\ along with the experimental points of Section~\ref{theory}. The
permittivity is calculated with (\ref{clamped}), since due to smallness of
the piezomodule $d_{14}$
\cite{a15} the difference between transverse permittivities of clamped and
free \dekdp\ can be neglected;  the pressure dependence of the dipole moment
$\mu_1$ is given by (\ref{effective}).

The theory qualitatively well reproduces the temperature curve of the
permittivity, including a decrease
with temperature in the paraelectric phase and the jump at the transition
point, but is not able to explain the existence of a small broad maximum of
$\eps_{11}$ at temperature slightly higher than the transition point. That
is a common drawback of all existing calculations of the transverse
permittivity, that can be removed by assuming a temperature dependent
high-frequency contribution to the permittivity $\eps_{1\infty}$.

As one can see, a satisfactory quantitative agreement with experimental
data for the rates of a decrease in the permittivity with pressure in the
high-temperature phase and of an increase in the low-temperature phase is
obtained. The fact that the value of the ratio
$\delta_1/\delta_0$, yielding the correct theoretical pressure dependence of
transition temperature, provides the correct pressure dependence of
the transverse dielectric permittivity, no extra fitting parameters
being introduced into the theory, gives one more evidence on the
important role played by the hydrogen bonds geometry, namely the separation
 $\delta$ between equilibrium deuteron sites, in the phase transition and
dielectric response of the hydrogen bonded crystals.

\section{Concluding remarks}
We performed experimental studies of hydrostatic pressure influence
on transverse and longitudinal dielectric permittivities of
ferroelectric \kdp\ and \dekdp\ crystals at 1MHz, the frequency that
belongs to the piezoelectric resonance region.

The obtained temperature curves of transverse permittivity $\eps_{11}$
are similar to the corresponding curves of static permittivity of
the crystals, while the pressure dependences of the permittivities are
analogous to those in the antiferroelectric crystals  \ada\ and \dada.

Unlike $\eps_{11}$, the temperature curves of the longitudinal permittivity
$\eps_{33}$
at 1MHz are qualitatively different from static ones. The multipeak
structure of the permittivity in the vicinity of the
transition point is observed. Under hydrostatic pressure, the two minima of
the longitudinal permittivity are getting closer, so that its two-minimum
structure transforms to a single-minimum one.
Above the transition point, the longitudinal permittivities of \kdp\ and
\dekdp\ obey the Curie-Weiss law; the pressure dependences of the
corresponding Curie constants well accord with the data for the Curie
constants of static permittivities.

The measured rates of decrease with pressure in the transition
temperature and permittivity $\eps_{11}$ in the paraelectric phase and of
a increase in $\eps_{11}$ in the ferroelectric phase in a deuterated \dekdp\
are well described within the proton ordering model. Theoretical
pressure dependences of $T_{\rm C}$ and $\eps_{11}$ to a great extent are
governed by the ratio $\delta_1/\delta_0$. This parameter denotes  relative
pressure changes in the separation $\delta$ between equilibrium deuteron
sites on a bond. Obtained in the present paper theoretical description of
the pressure dependence of  $\eps_{11}$, a similar agreement with
experimental data for spontaneous polarization and longitudinal static
dielectric permittivity \cite{our!}, as well as the previously revealed
universality of \cite{our!,Nel5} of the transition temperature vs
H(D)-site distance $\delta$ in the MeD$_2$XO$_4$ (Me = K, Rb, ND$_4$, X = P,
As), \kdp\ and \adp\ crystals, indicate that an important role in the phase
transition and  dielectric response of the crystals is played by the
geometry of hydrogen bonds, in particular, by the separation  $\delta$
between hydrogen sites.

So far we have no clear explanation of the multipeak structure of the
longitudinal permittivity. Apparently, it is a dynamic effect
connected with a  piezoelectric resonance phenomena, since at lower
frequencies as well as at frequencies above the  piezoelectric resonance
but below the dielectric dispersion region \cite{Hill} the  $\eps_{33}$
has a typical form of the permittivity with one peak at the  transition
point and the  Curie-Weiss behavior in the paraelectric phase.

The  maxima of $\eps_{33}$ in the ferroelectric phase might be merely the
peaks at different piezoelectric resonance frequencies (harmonics). The
harmonics frequencies are determined by sample dimensions and by the
appropriate elastic constants (only by  $c_{66}^E$ for a 45$^o$ Z cut
sample). The $c_{66}^E$ has a peculiarity at the phase transition point
\cite{Mason_old}, dropping from about $6\cdot 10^{10}$ dyn/cm$^2$ to zero
and increasing back to this value in the paraelectric phase. Therefore, at
changing temperature in the vicinity of the transition point, the resonant
harmonics frequencies span a wide frequency range, so that at certain
temperatures they may coincide with the measuring frequency 1~MHz.

\section*{Acknowledgments}
This work was supported by the Foundation for
Fundamental Investigations of the Ukrainian Ministry in Affairs of Science
and Technology, project No 2.04/171.

\end{document}